\documentclass[12pt,english]{article}
\usepackage{graphicx}
\usepackage{setspace}
\usepackage{amsmath}
\usepackage{url}
\usepackage[authoryear]{natbib}
\usepackage[verbose,a4paper,tmargin=2.54cm,bmargin=2.54cm,lmargin=2.54cm,rmargin=2.54cm]{geometry}
\usepackage{lineno} %command for numberling lines. It needs file ``lineno.sty''

\title{Nested species interactions promote feasibility over stability during the assembly of a pollinator community}

\author{Serguei Saavedra$^{1,2}$ \footnote{To whom correspondence should be addressed. E-mail: serguei.saavedra@usys.ethz.ch} \footnote{These authors contributed equally to this work} , Rudolf P. Rohr$^{3}$\footnotemark[\value{footnote}], \\
Jens M. Olesen$^4$, Jordi Bascompte$^{1}$\\
\vspace{0.09 in}
\\ 
$^{1}$Department of Evolutionary Biology and Environmental Studies \\University of Zurich \\ Winterthurerstrasse 190, CH-8057 Zurich, Switzerland\\ \\
$^{2}$Department of Environmental Systems Science, ETH Zurich, \\ Universit\"{a}tstrasse 16, CH-8092 Zurich, Switzerland\\ \\
$^{3}$Department of Biology - Ecology and Evolution, 
University of Fribourg \\ Chemin du Mus\'{e}e 10, CH-1700 Fribourg, Switzerland \\ \\
$^{4}$Department of Bioscience, Aarhus University, \\ Ny Munkegade 166, DK-8000 Aarhus, C Denmark
}
\date{
}

\begin{document}
\maketitle
\baselineskip=8.5 mm

\newpage
\begin{spacing}{1.9}
\raggedright

%%\begin{linenumbers}

\section*{Abstract}

The foundational concepts behind the persistence of ecological communities have been based on two ecological properties: dynamical stability and feasibility. The former is typically regarded as the capacity of a community to return to an original equilibrium state after a perturbation in species abundances, and is usually linked to the strength of interspecific interactions. The latter is the capacity to  sustain positive abundances on all its constituent species, and is linked to both interspecific interactions and species demographic characteristics. Over the last 40 years, theoretical research in ecology has emphasized the search for conditions leading to the dynamical stability of ecological communities, while the conditions leading to feasibility have been overlooked. However, thus far, we have no evidence of whether species interactions are more conditioned by the community's need to be stable or feasible. Here, we introduce novel quantitative methods and use empirical data to investigate the consequences of species interactions on the dynamical stability and feasibility of mutualistic communities. First, we demonstrate that the more nested the species interactions in a community are, the lower the mutualistic strength that the community can tolerate without losing dynamical stability. Second, we show that high feasibility in a community can be reached either with high mutualistic strength or with highly nested species interactions. Third, we find that during the assembly process of a seasonal pollinator community located at The Zackenberg Research Station (NE Greenland), a high feasibility is reached through the nested species interactions established between newcomer and resident species. Our findings imply that nested mutualistic communities promote feasibility over stability, which may suggest that the former can be key for community persistence.

\vspace{0.2 in}
{\sc Keywords: } coexistence, global stability, feasibility, mutualistic networks, mutualistic strength, nestedness

%\newpage

\section*{Introduction}

How can ecological communities sustain a large number of species? This is a major question that has greatly intrigued ecologists since the 1920's \citep{Elton27,Odum53,MacArthur55,Elton58,Margalef68}. Two ecological properties have been considered the foundational concepts behind the persistence of ecological communities: dynamical stability and feasibility \citep{MacArthur55,Vandermeer70,Gardner,May72,Roberts,Angelis,Vandermeer,Goh79,Yodzis,Svirezhev,Logofet}. Dynamical stability (hereafter stability) asks whether a community will return to an assumed equilibrium state after a perturbation in species abundances, and it is linked to the strength of interspecific interactions \citep{Svirezhev,Logofet}. Feasibility corresponds to the range of tolerated combinations of species demographic characteristics (intrinsic growth rates or carrying capacities) under which all species can have positive abundances \citep{Vandermeer70,Vandermeer,Svirezhev,Logofet,Bastolla,Nattrass12,Rohr,SaaRohr}. Importantly, the conditions leading to the stability of a community does not automatically imply its feasibility and vice versa \citep{Vandermeer70,Roberts,Vandermeer,Svirezhev,Stone,Logofet,Rohr}.

\vspace{0.25 in}

Over the last 40 years, theoretical research in ecology has emphasized the search for conditions leading to the stability of ecological communities \citep{May72,Angelis,Goh79,Yodzis,Svirezhev,Logofet,Staniczenko}, while the conditions leading to feasibility have received considerably less attention \citep{Vandermeer70,Roberts,Vandermeer,Svirezhev,Logofet,Hofbauer,Nattrass12,Rohr}. However, theoretical and empirical studies have shown that the sequence of community assembly cannot be understood without feasibility conditions \citep{Drake,Law,Weatherby,Saavedra2009c}. Yet, it is still unclear the extent to which species interactions are more conditioned by the community's need to be stable or feasible. This is important in order to better understand the link between community structure and dynamics, especially as global environmental change is accelerating the rate at which species are removed and introduced into new habitats \citep{Sala,Tylianakis}.

\vspace{0.25 in}

To answer the above question, we introduce general quantitative methods to investigate the role of stability and feasibility in shaping mutualistic communities. Because stability and feasibility are linked and conditioned by species interactions, we study the general association between stability and feasibility in mutualistic communities, and how this association is modulated by species interaction networks. We then move to an empirical case by studying how the association between stability and feasibility acts on the assembly of a seasonal Arctic pollinator community located at The Zackenberg Research Station, Northeastern Greenland (748300 N, 218000 W). Finally, we discuss the implications of our findings.

\section*{Methods}

\subsection*{Mutualistic model}

To study the conditions compatible with stability and feasibility in mutualistic communities, we used a generalized Lotka-Volterra model of the form:

\begin{equation*} \label{equ:ode}
\left\lbrace \begin{array}{c}
	\frac{dP_{i}}{dt} = P_{i} \Big( r_{i}^{(P)} - \sum_{j} \alpha_{ij}^{(P)} P_j + \sum_{j} \gamma_{ij}^{(P)} A_{j} \Big) \\
	\frac{dA_{i}}{dt} = A_{i} \Big( r_{ik}^{(A)} - \sum_{j} \alpha_{ij}^{(A)} A_j + \sum_{j} \gamma_{ij}^{(A)} P_{j} \Big)
\end{array} \right.,
\end{equation*} 
where the variables $P_{i}$ and $A_{i}$ denote the abundance of plant and animal species $i$, respectively. The parameters of this mutualistic model correspond to the values describing intrinsic growth rates ($r_{i}$), intra-guild competition ($\alpha_{ij}$), and the benefit received via mutualistic interactions ($\gamma_{ij}$). The mutualistic benefit is parameterized by $\gamma_{ij} = \big( \gamma_{0} y_{ij} \big) / \big(d_{i}^{\delta} \big)$, where $y_{ij} = 1$ if species $i$ and $j$ interact and zero otherwise; $d_{i}$ is the number of interactions of species $i$; $\delta$ corresponds to the mutualistic trade-off \citep{Saavedra}; and $\gamma_0$ represents the overall level of mutualistic strength. We used a mean field approximation for the competition parameters, where we set $\alpha_{ii}^{(P)} = \alpha_{ii}^{(A)} = 1$ and $\alpha_{ij}^{(P)} = \alpha_{ij}^{(A)} = \rho$ ($i \neq j$). We analyzed two important cases of this model, where the interspecific competition is zero ($\rho=0$, $0<\delta<0.5$) and where the mutualistic trade-off is zero ($0<\rho<0.01$, $\delta = 0$). These two extreme cases allowed us to focus on the effects of mutualistic interactions (the cornerstone of pollinator networks) on the conditions for species coexistence. We used a linear version of a mutualistic model (i.e., there is no density saturation as the strength of mutualism increases) because, as opposed to a nonlinear version, results can be analytically tractable \citep{Rohr}. Importantly, under the explored parameterization, the dynamical behavior of the community remains general in a nonlinear version of this model \citep{Saavedra,Rohr}. Further details are given below.

\subsection*{Stability conditions}

Traditionally, the stability of a community has been investigated by looking at its local asymptotic stability \citep{May72}. This type of stability asks whether a community will return to an equilibrium point after an infinitesimal small perturbation in species abundances. An equilibrium point is the state of species abundances ($A^*_{i}$ and $P^*_{i}$) at which the community does not change anymore. Importantly, for the studied mutualistic model, it is possible to conclude more than only the local behavior of the abundance trajectories of a community. For instance, if certain conditions are satisfied, the community can have only one globally stable equilibrium point, meaning that all the abundance trajectories, regardless of their initial position (as long as the initial abundances are strictly positive), converge to that unique equilibrium point. Otherwise, the community may have alternative stable states or unbounded abundance trajectories.

\vspace{0.25 in}

In particular, in the studied mutualistic model, the conditions determining the convergence of the community to only one globally stable equilibrium point depend on the interaction strength matrix $\boldsymbol{\alpha}=\begin{bmatrix} \boldsymbol{\alpha^{(P)}} & \boldsymbol{-\gamma^{(P)}}  \\ \boldsymbol{-\gamma^{(A)}} & \boldsymbol{\alpha^{(A)}} \end{bmatrix}$, embedding all the competition ($\alpha_{ij}$) and mutualistic ($\gamma_{ij}$) interactions between species $i$ and $j$ \citep{Rohr}. Importantly, within our parameterization of interaction strengths, as long as the real parts of all eigenvalues of $\boldsymbol{\alpha}$ are positive, the matrix is Volterra dissipative and, therefore, there exist only one globally stable equilibrium point \citep{Volterra,Takeuchi78,Goh79,Svirezhev,Logofet,Hofbauer,Rohr}. In the case where the interspecific competition is zero ($\rho=0$, $0<\delta<0.5$), the proof of global stability is achieved by using M-matrix theory \citep{Goh79,Svirezhev,Logofet,Hofbauer}, while in the case where the mutualistic trade-off is zero ($0<\rho<0.01$, $\delta = 0$), the proof is based on the symmetry of the interaction matrix \citep{Svirezhev,Logofet,Rohr}.

\vspace{0.25 in}

The above statement implies that in the studied model there is an upper limit to the mean mutualistic strength in the community $\langle \gamma_{ij} \rangle$ below which
the abundance trajectories converge to a unique globally stable equilibrium point \citep{Goh79,Rohr}. This upper limit $\hat{\gamma}$ can be calculated by finding the mean mutualistic strength $\langle \gamma_{ij} \rangle$ at which the real part of one of the eigenvalues of the corresponding matrix $\boldsymbol{\alpha}$ reaches zero (see Fig. 1A). This upper limit of tolerated mutualistic strength is what we called the stability condition. The higher the value of $\hat{\gamma}$, the larger the mutualistic strength that can be tolerated in the community without losing global stability. Importantly, this upper limit is approximately equivalent to the upper limit conditioning the global stability in the nonlinear version of the model \citep{Goh79,Saavedra,Rohr}. 

\vspace{0.25 in}

Unfortunately, even global stability does not guarantee that all species will survive in the community \citep{Vandermeer70,Roberts,Svirezhev,Logofet,Rohr,Saavedra15}. At the unique equilibrium point, some species may have an abundance of zero. This means the the abundance trajectories of the dynamical system can go towards the border (i.e., where at least one of the species abundances goes to zero) and consequently some species go extinct. To have survival of all species, we need a second condition to constrain that at the unique globally stable equilibrium point all species have strictly positive abundances. This second condition is called feasibility \citep{Vandermeer70,Roberts,Svirezhev,Logofet,Hofbauer,Rohr,Saavedra15}.

\subsection*{Feasibility conditions}

The feasibility of an equilibrium point corresponds to the conditions leading to positive species abundances ($A^*_{i} >0$ and $P^*_{i}>0$). As opposed to stability, the conditions determining feasibility are determined by both the interaction strength matrix $\boldsymbol{\alpha}$ and the vector of intrinsic growth rates $\boldsymbol{r}$ \citep{Rohr}. Therefore, if the interaction strength matrix of a community is known, the feasibility of that community only depends on the domain of intrinsic growth rate vectors $\boldsymbol{r}$ leading to positive abundances. This implies that a community can be globally stable by virtue of its interaction strength matrix $\boldsymbol{\alpha}$, but either feasible  ($A^*_{i} >0$ and $P^*_{i}>0$) or not (one or more of the species abundances are equal or lower than zero) depending on the vector of intrinsic growth rates $\boldsymbol{r}$.

\vspace{0.25 in}

As shown in Figure 1B, the domain of intrinsic growth rate vectors leading to positive species abundances is geometrically described by an algebraic cone, where the borders are established by the column vectors of the corresponding interaction strength matrix $\boldsymbol{\alpha}$ \citep{Svirezhev,Logofet,Saavedra15}. The solid angle of that cone ($\Omega$) generated by the matrix $\boldsymbol{\alpha}$ can be interpreted (given the right normalization) as the probability of sampling randomly a vector of intrinsic growth rates that fall inside that cone \citep{Svirezhev,Logofet,Saavedra15}. The normalization can be done without loss of generalization by sampling the vectors of intrinsic growth rates uniformly on the sphere using the following integration \citep{Ribando}:

\begin{equation*}\label{Omega}
	{\Omega} = \frac{|\det(\boldsymbol{\alpha})|}{\pi^{S/2}} \idotsint_{\boldsymbol{R}^S_{\geq 0}}  e^{-\boldsymbol{x}^T \boldsymbol{\alpha}^T \boldsymbol{\alpha} \boldsymbol{x} }d\boldsymbol{x}.
\end{equation*}

Moreover, by setting $\boldsymbol{\alpha}^T \boldsymbol{\alpha} = \frac{1}{2} \Sigma^{-1}$, the above integration transforms into: 
\begin{equation*}\label{Omega2}
	{\Omega} = \frac{1}{(2\pi)^{S/2} \sqrt{|\det(\Sigma)|}} \idotsint_{\boldsymbol{R}^S_{\geq 0}}  e^{-\boldsymbol{x}^T  \frac{1}{2} \Sigma^{-1} \boldsymbol{x} }d\boldsymbol{x},
\end{equation*}
which is then the cumulative distribution function of a multivariate normal distribution of mean zero and variance-covariance matrix $\Sigma$. Such a function can be evaluated efficiently by a quasi-Monte Carlo method \citep{Genz,mvtnorm}. This normalized solid angle ${\Omega}$ is what we called the feasibility condition (i.e., the domain of intrinsic growth rates leading to positive abundances for all species). ). This solid angle is computed and represented on a log scale. The higher the value of $\Omega$, the larger the likelihood of finding feasibility in the system \citep{Svirezhev,Logofet,Saavedra15}. In general, the size of the cone increases with the overall level of mutualism \citep{Rohr}.

\vspace{0.25 in}

Finally, it is worth noting that, in our model, if at least one eigenvalue derived from the interaction strength matrix has negative real part, then even if there exists an equilibrium point with strictly positive abundances for all species, this feasible equilibrium point is unstable. In that case, the abundance trajectories of the dynamical system will go towards the border, and at least one of the species will eventually go extinct \citep{Goh79,Hofbauer}. Only when both the stability and feasibility conditions are satisfied, the abundance trajectories will not go towards the border, and will allow species coexistence in the long run.

\subsection*{Nestedness}

As mentioned before, stability and feasibility are linked to and conditioned by species interactions, specifically by the mutualistic interaction strengths $\gamma_{ij} = \big( \gamma_{0} y_{ij} \big) / \big(d_{i}^{\delta} \big)$. Therefore, it becomes necessary to understand how these species interaction networks impact both stability and feasibility. Research in mutualistic networks has shown that a highly nested pattern of interactions can minimize the competitive effects between species \citep{Bastolla}, minimize local stability \citep{Staniczenko}, and increase the likelihood of community persistence \citep{Rohr}. A highly nested pattern can be equivalent to a high fraction of shared interactions between species \citep{Bascompte03,Bastolla}. Therefore, nestedness gives a description of species interaction networks that can be linked to community dynamics. Following \cite{Bastolla}, the nestedness of a network can be calculated as:

\begin{equation*}
n=\frac{\sum_{i<j}d^{(A)}_{ij} + \sum_{i<j}d^{(P)}_{ij}}{\sum_{i<j} \text{min}(d^{(A)}_{i},d^{(A)}_{j}) + \sum_{i<j} \text{min}(d^{(P)}_{i},d^{(P)}_{j})},
\end{equation*}
where $d_{ij}$ corresponds to the number of shared interactions between species $i$ and $j$, $d_{i}$ corresponds to the number of interactions of species $i$, the variables $A$ and $P$ correspond respectively to animals and plants, and $\text{min}(d^{(P)}_{i},d^{(P)}_{j})$ refers to the smallest of the two values. This measure takes values between 0 and 1, where higher the values, the higher the nestedness of a species interaction network. 

\section*{Results}

\subsection*{General case}

We studied the general association between stability and feasibility, and how this association is modulated by species interaction networks---summarized by nestedness. To carry out this analysis, we generated several mutualistic networks with the same number of animals, plants, and interactions. In each of these networks, species interactions were established randomly between animals and plants, such that we generate a broad gradient of nestedness values with the same number of species and interactions. For each generated network $g$, we calculated its stability condition ($\hat{\gamma}^g$), feasibility condition ($\Omega^g$), and level of nestedness ($n^g$). Stability and nestedness only depend on the generated network $g$, while feasibility is calculated over the interaction strength matrix $\boldsymbol{\alpha}$, which depends on both the generated network and the mean mutualistic strength $\langle \gamma_{ij} \rangle$. Therefore, we studied how feasibility changes as a function of both nestedness and the mean mutualistic strength.

\vspace{0.25 in}

First, we find that nestedness is strongly and negatively associated with stability. Figures 2A-B show that for a community with 17 plants, 24 animals, and 140 interactions, with and without interspecific competition, the maximum mutualistic strength $\hat{\gamma}^g$ that the community can tolerate without losing stability decreases with the level of nestedness $n^g$ (Spearman rank correlations of $r<-0.98$, $P<10^{-3}$). Note that the range of nestedness values is lower than the theoretical range described between 0 and 1. This is because the actual minimum and maximum nestedness values in a network are constrained by the number of species and interactions \citep{Rohr}.

\vspace{0.25 in}

Second, we find that high feasibility in a community can be reached either with high mutualistic strength or with highly nested species interactions. For the same community used before, Figures 2C-D show the relationship between nestedness and feasibility for high, medium and low values of mean mutualistic strength (the darker the symbol, the higher the mutualistic strength). Note that the higher the mean mutualistic strength is, the smaller the fraction of generated networks that can tolerate that strength without losing stability. Importantly, the figures show that high feasibility values can be reached by generated networks with either a high mutualistic strength and low nestedness, or a low mutualistic strength and high nestedness. Indeed, comparing generated networks with the same mean mutualistic strength (dashed lines), there is a strong and positive correlation between nestedness and feasibility (Spearman rank correlations of $r>0.88$, $P<10^{-3}$). This pattern was found in any given community with any combination of number of species and interactions under the explored parameterization. This reveals a community trade-off between stability and feasibility tuned by the nested architecture of species interactions. 

\vspace{0.25 in}

Importantly, these findings imply that nested mutualistic communities promote feasibility over stability. Therefore, a question that remains to be answered is whether observed species interactions are conditioned by the community's need to be stable or feasible. In other words, are communities reaching a high feasibility? And if so, are communities reaching this through a high mutualistic strength or through highly nested interactions?

\subsection*{Empirical case}

To answer the above questions, we used empirical data describing the assembly process of an Arctic pollinator community located at The Zackenberg Research Station, Northeastern Greenland (748300 N, 218000 W). In this community, day by day newcomer species (both flowering plants and pollinators) join the resident species according to their own phenophase. Along the observation period, the community experiences an increase and decrease in the number of species and interactions from the last snow melted to the first snowfall in the site (see Fig. 3). Our study period covers two full seasons (1996 and 1997), where observations were recorded daily whenever weather conditions allowed. From a 3-month period in each season, bad weather reduced the number of observation days to 23 and 25 for 1996 and 1997, respectively. For each day, our data record the identity of resident species leaving the community, the identity of newcomer species joining the community, and the new established interactions between newcomer and resident species (the data is provided as Supplementary Information). See \cite{Olesen08} for full details about the data and study site. 

\vspace{0.25 in}

For each day, to calculate the extent to which the interactions established by the newcomer species modulate the stability and feasibility conditions, we explored all the different network combinations that could be established by rewiring the interactions between newcomer and resident species. Because phenophase length has been reported as an important correlate of the assembly process \citep{Olesen08}, our rewiring procedure always preserves the identity and number of species observed in each day. This rewiring procedure also keeps both the observed number of interactions per day and the interactions between resident species. For each day $k$, we calculated the corresponding stability ($\hat{\gamma}_k^g$), feasibility ($\Omega_k^g$), and nestedness ($n_k^g$) from each rewired network $g$. To calculate $\Omega_k^g$, we used a fixed value of $\langle \gamma_{ij} \rangle _k ^g= \hat{\gamma}_k /2$, where $\hat{\gamma}_k$ corresponds to the maximum tolerated mutualistic strength of the observed network in day $k$. This fixed value was chosen so that all rewired networks can be stable. This value does not change qualitatively our results as long as the networks are stable.

\vspace{0.25 in}

In line with our general results, we find that despite the constraints imposed by both species phenophase and the limited number of interactions between newcomer and resident species that can be rewired, daily interaction networks also show a trade-off between stability and feasibility tuned by the nested architecture. Figure 4 shows the Spearman rank correlations between nestedness and stability (open squares) as well as between nestedness and feasibility (solid circles). The figure shows that in both years and taking into account or not interspecific competition, nestedness ($n_k^g$) is always strongly negatively and strongly positively correlated with stability ($\hat{\gamma}_k^g$) and feasibility ($\Omega_k^g$), respectively. This implies that newcomer species through their established mutualistic interactions can promote either stability or feasibility during the assembly process, but not both at the same time.

\vspace{0.25 in}

To investigate the extent to which newcomer species promote feasibility in each day $k$, we investigated the maximum level of feasibility $\hat{\Omega}_k^g$ that can be reached by any given rewired network $g$ and compared it to the maximum feasibility $\hat{\Omega}_k$ that can be reached by the observed network in each day. Because feasibility increases with mutualistic strength, the maximum feasibility in each network was calculated using the maximum tolerated mutualistic strength ($\hat{\gamma}_k^g$). The comparison then was evaluated using the scaled maximum feasibility $\Omega_k^s=(\hat{\Omega}_k-\min(\hat{\Omega}_k^g))/(\max(\hat{\Omega}_k^g)-\min(\hat{\Omega}_k^g))$, where  $\hat{\Omega}_k$ is the observed maximum feasibility condition in day $k$, and $\max(\hat{\Omega}_k^g)$ and $\min(\hat{\Omega}_k^g)$ are the maximum and minimum values of the maximum feasibility conditions found in the rewired networks in day $k$, respectively. These scaled values range between 0 and 1, where higher the values, the more the observed interactions established by newcomer species approach the maximum possible feasibility conditions that can be reached by the community in a given day. As these scaled values explicitly consider each possible rewiring scenario, they have advantages over previous relative measures (e.g., $p$-values) that are sensitive to the specific choice of null model and community size \citep{Saavedra13}. Our results are qualitative the same when using ranked values instead of scaled values (i.e., ranked position of the observed value within the generated values), which confirms that our results are also robust to the variance in the distribution of values.

\vspace{0.25 in}

We find that in both years and taking into account or not interspecific competition, the observed species interactions can reach feasibility conditions that are close to the maximum possible in each day. Figure 5 shows that the majority of scaled feasibility values (88 out of 96, binomial test $P<10^{-3}$) have values larger than 0.5, revealing that the observed interactions established by newcomer species are lying in the upper half of the potential range of feasibility conditions in any given day.

\vspace{0.25 in}

Finally, to see whether these high feasibility conditions are due to a high mutualistic strength or nested species interactions, we investigated the extent to which the maximum tolerated mutualistic strength and nestedness is promote by the observed networks in very single day. The analysis was carried out using the scaled stability and scaled nestedness values. The scaled stability is calculated by $\gamma_k^s=(\hat{\gamma}_k-\min(\hat{\gamma}_k^g))/(\max(\hat{\gamma}_k^g)-\min(\hat{\gamma}_k^g))$, where $\hat{\gamma}_k$ is the observed maximum tolerated mutualistic strength in day $k$, and $\max(\hat{\gamma}_k^g)$ and $\min(\hat{\gamma}_k^g)$ are the maximum and minimum values of maximum tolerated mutualistic strength found in the rewired networks $g$ in day $k$, respectively. Similarly, the scaled nestedness is calculated by $n_k^s=(n_k-\min(n_k^g))/(\max(n_k^g)-\min(n_k^g))$, where $n_k$ is the observed nestedness value in day $k$, and $\max(n_k^g)$ and $\min(n_k^g)$ are the maximum and minimum values of nestedness found in the rewired networks $g$ in day $k$, respectively. Again, these scaled values take values between 0 and 1, where higher the values, the more the observed interactions established by newcomer species approach the maximum possible mutualistic strength (nestedness) that can be reached by the community in a given day. Therefore, the higher (lower) the scaled stability values are relative to the scaled nestedness values, the more (less) the scaled feasibility value depends on the mutualistic strength and less (more) on the nested species interactions.

\vspace{0.25 in}

We find that in these communities, feasibility does not depend on mutualistic strength as much as it does on the nested species interactions. Figure 6 shows that in both years and taking into account or not interspecific competition, the majority of observed days (80 out of 96, paired $t$-test $P<10^{-3}$) the scaled nestedness values (closed triangles) are larger than the scaled stability values (open squares). Moreover, the figure shows that the majority of the scaled nestedness values (90 out of 96, binomial test $P<10^{-3}$) are larger than 0.5, revealing that the observed interactions established by newcomer species are lying in the upper half of the potential range of nestedness values in any given day. In contrast, the figure also shows that the minority of the scaled stability values (20 out of 96, binomial test $P<10^{-3}$) are larger than 0.5, revealing that the observed interactions established by newcomer species are lying in the lower half of the potential range of stability values in any given day. Importantly, these findings confirm that within the assembly possibilities of the observed mutualistic community, feasibility is promoted over stability, and this is linked to the nested species interactions established between newcomer and resident species.

\section*{Discussion} 

The above findings have a series of interesting implications. First, the fact that nestedness tunes a trade-off between feasibility and stability may imply that different ecosystem services in mutualistic systems are not in same direction \citep{Loreau,Turnbull}. This means that it is not guaranteed that one component of community dynamics could always be used as a proxy for another component.  While previous studies have emphasized the high level of nestedness in mutualistic communities, less attention has been given to why observed nestedness is not even higher. Our result on the trade-off between stability and feasibility may explain why there might be a limit to nestedness: a further increase of an already high feasibility can be counterbalanced by a strong decrease in stability.

\vspace{0.25 in}

Second, the finding that feasibility is increased via nested---as opposed as through an increase in mutualistic strength---in the empirical community may be explained by dynamical and biological constraints. The dynamical constraints may be imposed by the theoretical observation that high mutualistic strengths can push the community to shift from a weak to a strong mutualistic regime, which can easily take the community to rather unpredictable dynamics \citep{Bastolla,Saavedra,Rohr}. The biological constraints may originate from the empirical observation that mutualisms among free living species are of low specificity, which is compatible with the combination of coevolutionary convergence and complementarity \citep{Thompson05}. In both cases, communities, especially under short-term dynamics, may have a higher opportunity to increase feasibility by changing the organization of their interactions rather than by increasing the overall mutualistic strength.

\vspace{0.25 in}

Third, the finding that feasibility is being promoted over stability may confirm that under short-term dynamics, the community may not need to be highly dynamically stable in order for species to coexist. For instance, other studies have suggested that asynchronous dynamics, reducing the amplification of perturbations, or reducing the variability of the total abundance may have more biological relevance for the community than the capacity to return to an equilibrium point \citep{Loreau}. Importantly, these findings reveal that feasibility is an important condition for species coexistence even under short-term dynamics and requires further exploration.

\vspace{0.25 in}

Finally, it is noteworthy that over more than 40 years, many studies in theoretical ecology have been focused on the dynamical stability of ecological communities. In particular, on local asymptotic stability. Indeed, one of the long-standing questions in ecology has been whether large ecological communities will be more locally stable \citep{May72}. However, empirically and theoretically, there has been no evidence demonstrating that dynamical stability should be the most important ecological property leading to community persistence. In fact, our results show that dynamical stability might not be as relevant as feasibility for species coexistence in seasonal communities. This calls for a stronger research program on the factors modulating feasibility and alternative stability conditions in species interaction networks, as they can hold the key for a general theory of community persistence.\\

\newpage

{\bf Acknowledgments} 
Funding was provided by the European Research Council through an Advanced Grant (JB).\\
\vspace{0.5 in}
{\bf Competing financial interests} The authors declare no competing financial interests.\\
\vspace{0.5 in}
{\bf Author contributions} SS and RPR designed the study and performed the analysis; JB supervised the work; JMO provided the data, SS and RPR wrote a first draft of the manuscript; all authors contributed to revisions.\\
\vspace{0.5 in}
{\bf Data accessibility} 
Data and code in R software are provided as supplemental information (Dryad Digital Repository. http://dx.doi.org/0.5061/dryad.3pk73).
\\

%{\bf Competing financial interests} The authors declare no competing financial interests.
%\\

%{\bf Author contributions} All authors designed the research. MAF and NS compiled the data. SS and RPR designed the methods and performed the analysis. All authors analyzed the data and wrote the paper.

\pagebreak

\bibliography{bibliography}{}
\bibliographystyle{besjournals}

%\pagebreak

%\section*{Figure Captions}

%\vspace{0.5 in}

%Figure 1: 

%\vspace{1.5 in}

%Figure 2: 

%\vspace{1.0 in}

%Figure 3: 

%\end{linenumbers}
\end{spacing}

\begin{spacing}{1.9}
\begin{linenumbers}

\clearpage

\setcounter{figure}{0}

\begin{figure}[ht]
\centerline{\includegraphics*[width= 1.0 \linewidth]{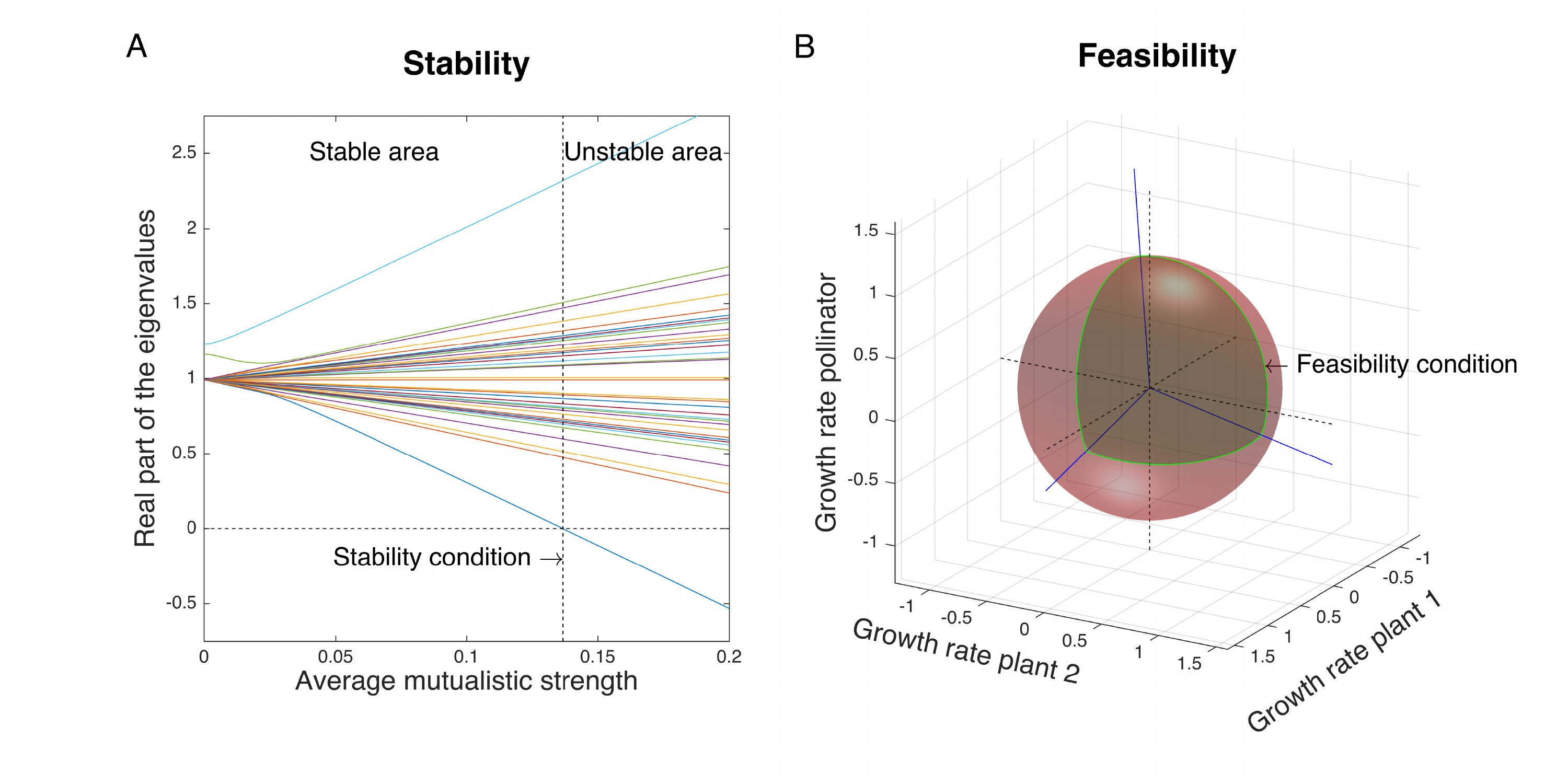}}
	\caption{Stability and feasibility conditions. Panel A shows the real part of each of the eigenvalues (lines) of the interaction strength matrix $\boldsymbol{\alpha}$ of a randomly generated community (with 24 animals, 17 plants, and 140 interactions) as a function of the mean mutualistic strength $\langle \gamma_{ij} \rangle$. The point at which one of the eigenvalues is lower or equal to zero (dashed line) becomes the maximum level of mutualistic strength at which the community can be globally stable. This point is what we called the stability condition $\hat{\gamma}$. The larger the value of $\hat{\gamma}$ is, the large the stability of the community. Panel B shows an illustration of the algebraic cone of feasibility for 3 species (two plants and one pollinator). The coordinates correspond to the intrinsic growth rates of the species. The cone (dark region) represents the hypervolume under which the community can sustain positive abundances for all species. This hypervolume is delimited by the column vector of the interaction strength matrix $\boldsymbol{\alpha}$ (blue solid lines). The sphere corresponds to the normalization of the cone. The normalized size of the cone (i.e., relative to the sphere) is what we called the feasibility condition $\Omega$. The larger the value of $\Omega$ is, the large the feasibility of the community.}
\label{fig1}
\end{figure}

\clearpage

\begin{figure}[ht]
\centerline{\includegraphics*[width= 1.0 \linewidth]{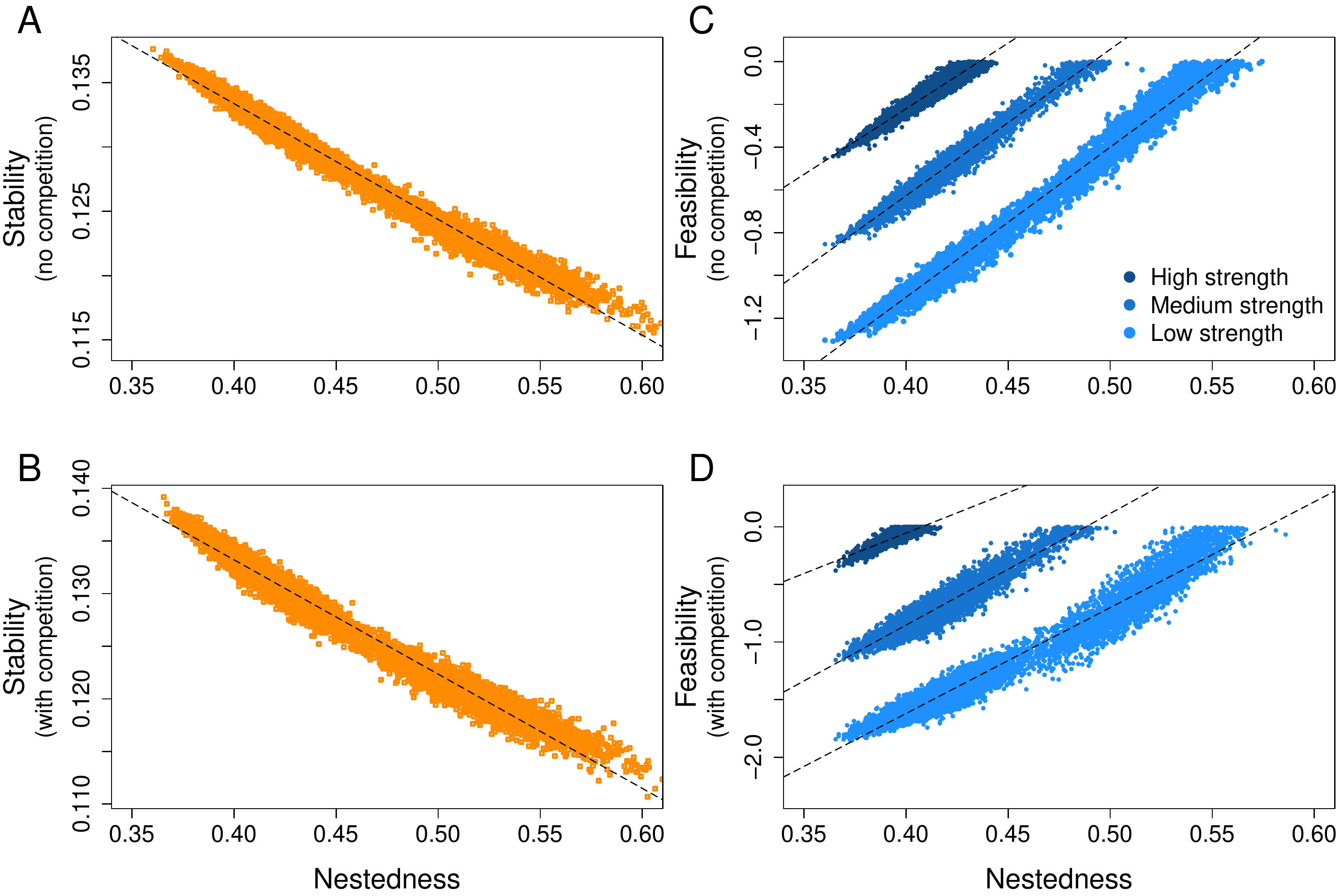}}
	\caption{Nestedness tunes the trade-off between stability and feasibility. Panels A and B show the negative association between nestedness ($n^g$) and stability (maximum tolerated mean mutualistic strength $\hat{\gamma}^g$) without and with interspecific competition for all the random generated networks $g$. Panels C and D show the positive association between nestedness ($n^g$) and feasibility ($\Omega^g$) without and with interspecific competition. The darker the symbol, the higher the mean mutualistic strength $\langle \gamma_{ij} \rangle$ used to calculate feasibility. For no competition, we used $\langle \gamma_{ij} \rangle=0.13$, $\langle \gamma_{ij} \rangle=0.125$, and $\langle \gamma_{ij} \rangle=0.12$ for a high, medium, and low mutualistic strength, respectively. For competition, we used $\langle \gamma_{ij} \rangle=0.133$, $\langle \gamma_{ij} \rangle=0.124$, and $\langle \gamma_{ij} \rangle=0.117$ for a high, medium, and low mutualistic strength, respectively. Each symbol corresponds to a generated network $g$ with 17 plants, 24 animals, and 140 species interactions. Interactions are randomly established in each generated network. The top panels correspond to the scenario with no interspecific competition ($\rho=0$ and $\delta=0.25$), and the bottom panels correspond to the scenario with interspecific competition ($\rho=0.01$ and $\delta=0$). All the other explored combinations of parameter values yield the same qualitative results. Note that the feasibility values are on a log scale.}
\label{fig2}
\end{figure}

\clearpage

\begin{figure}[ht]
\centerline{\includegraphics*[width= 1.0 \linewidth]{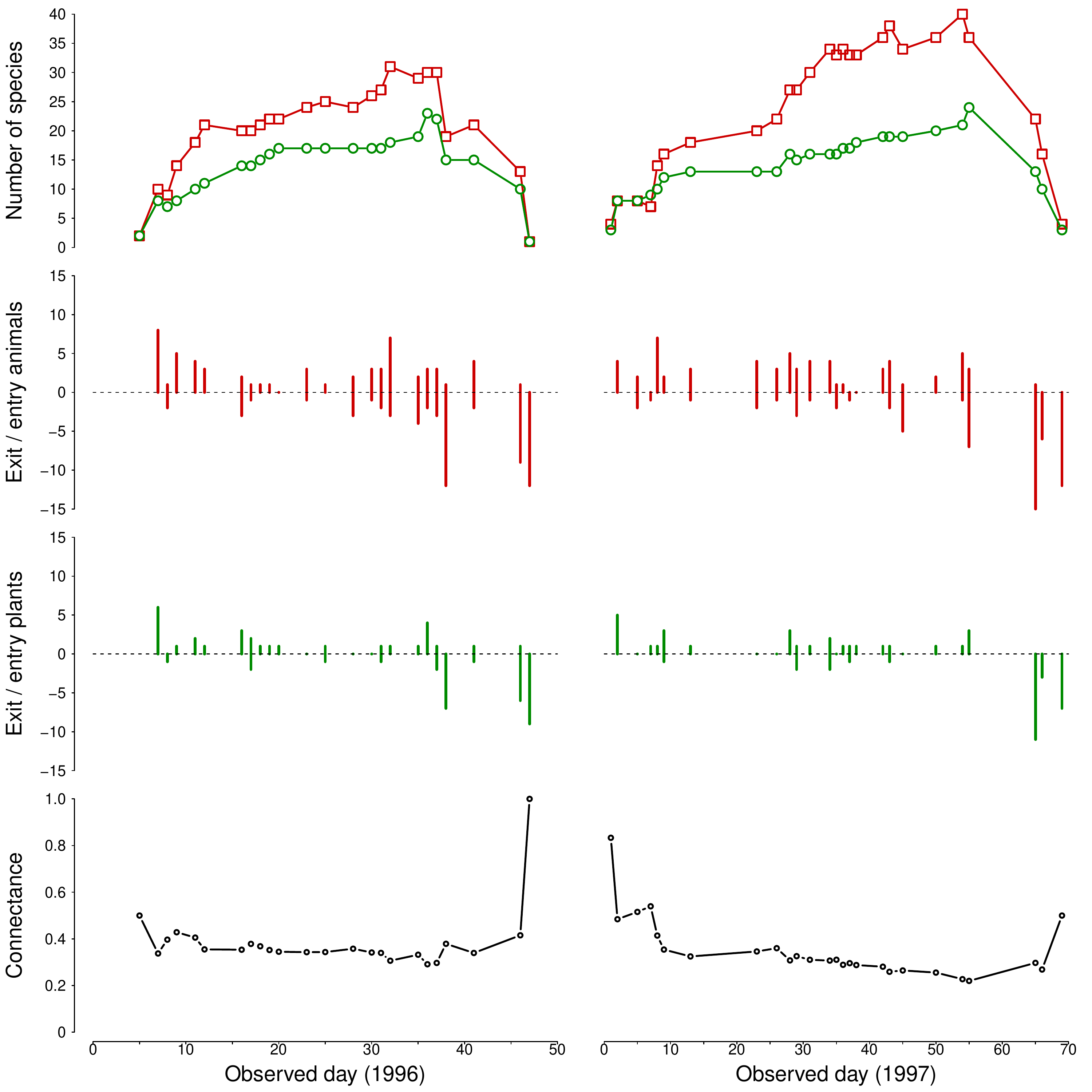}}
	\caption{Temporal dynamics of the observed pollinator network. The top panels illustrate the total number of animals and plants (red squares and green circles, respectively) at each observed day across the two observation periods (1996 and 1997). The middle panels correspond to the number of newcomer species (positive numbers) and resident species that exit the community (negative numbers) across the observation periods. The bottom panels correspond to the observed connectance in each day. Connectance is defined as the number of species interactions divided by the product of number of animals and plants.}
\label{fig3}
\end{figure}

\pagebreak

\begin{figure}[ht]
\centerline{\includegraphics*[width= 1.1 \linewidth]{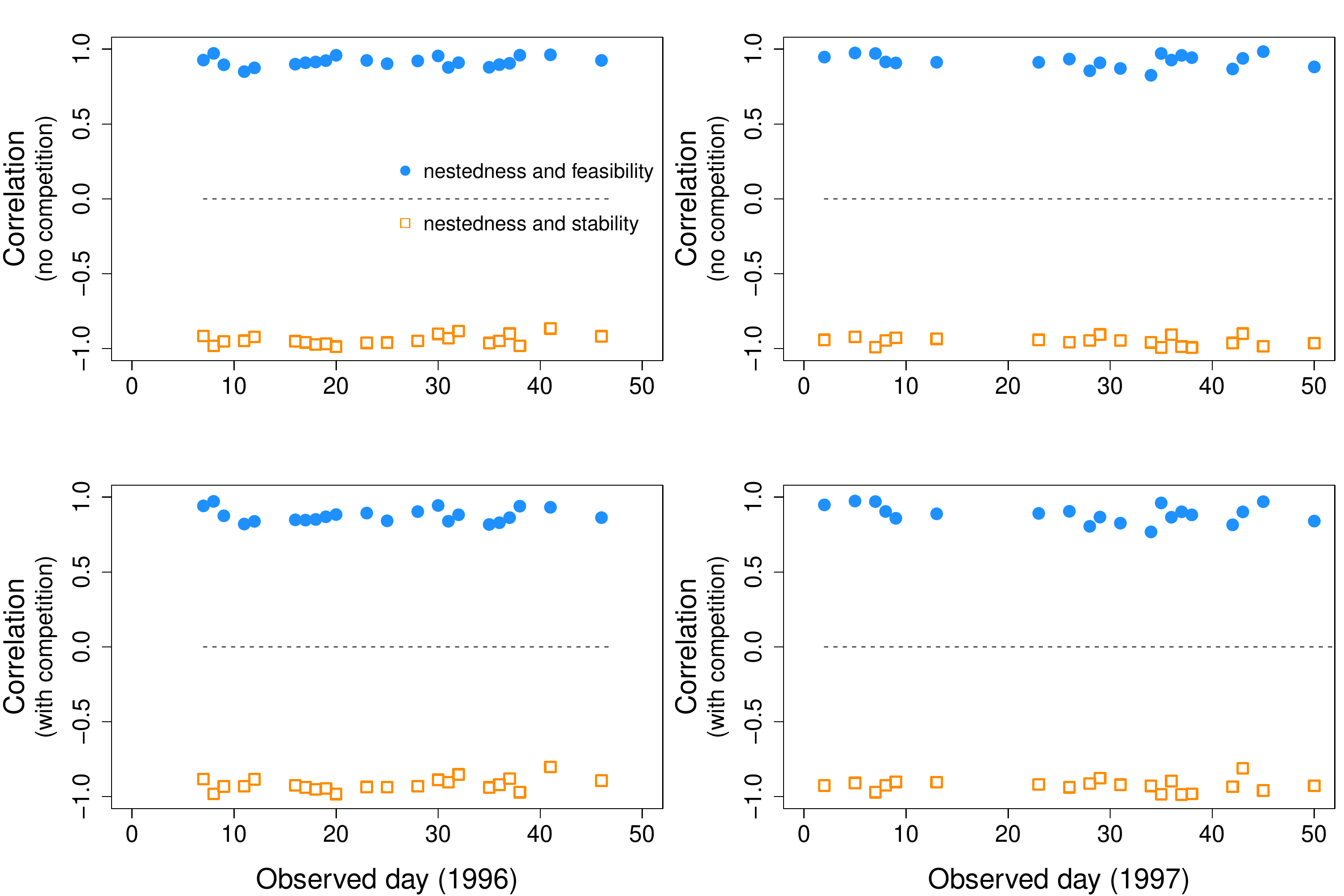}}
	\caption{Association of nestedness with stability and feasibility in rewired networks. The figure shows the Spearman rank correlations between nestedness $n_k^g$ and stability $\hat{\gamma}_k^g$ (open squares) and nestedness $n_k^g$ and feasibility $\Omega_k^g$ (closed circles). Each symbol corresponds to the correlation observed in each day $k$ across the two periods (1996 and 1997). These correlations are extracted from the rewired networks $g$ of the empirical communities in each day. The top panels correspond to the scenario with no interspecific competition ($\rho=0$ and $\delta=0.25$), and the bottom panels correspond to the scenario with interspecific competition ($\rho=0.01$ and $\delta=0$). All the other explored combinations of parameter values yield the same qualitative results.}
\label{fig4}
\end{figure}

\begin{figure}[ht]
\centerline{\includegraphics*[width= 1.2 \linewidth]{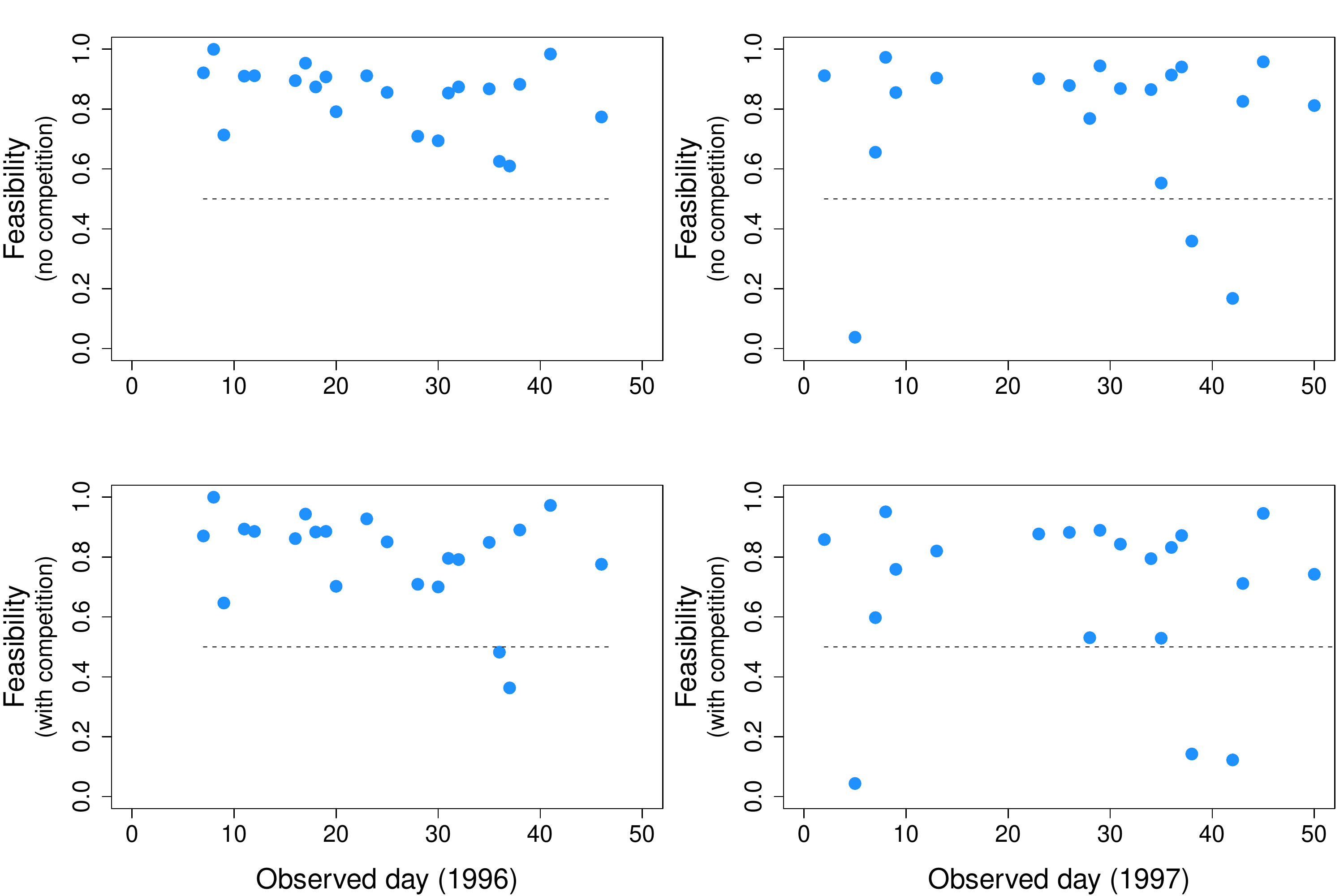}}
	\caption{Species interactions promote feasibility. The figure shows the scaled feasibility values in each of the observed days across the two periods (1996 and 1997). Each symbol corresponds to the scaled value in a day, and represents the position of the empirical network within the range of values generated from the rewired networks $g$. The top panels correspond to the scenario with no interspecific competition ($\rho=0$ and $\delta=0.25$), and the bottom panels correspond to the scenario with interspecific competition ($\rho=0.01$ and $\delta=0$). All the other explored combinations of parameter values yield the same qualitative results.}
\label{fig5}
\end{figure}

\begin{figure}[ht]
\centerline{\includegraphics*[width= 1.2 \linewidth]{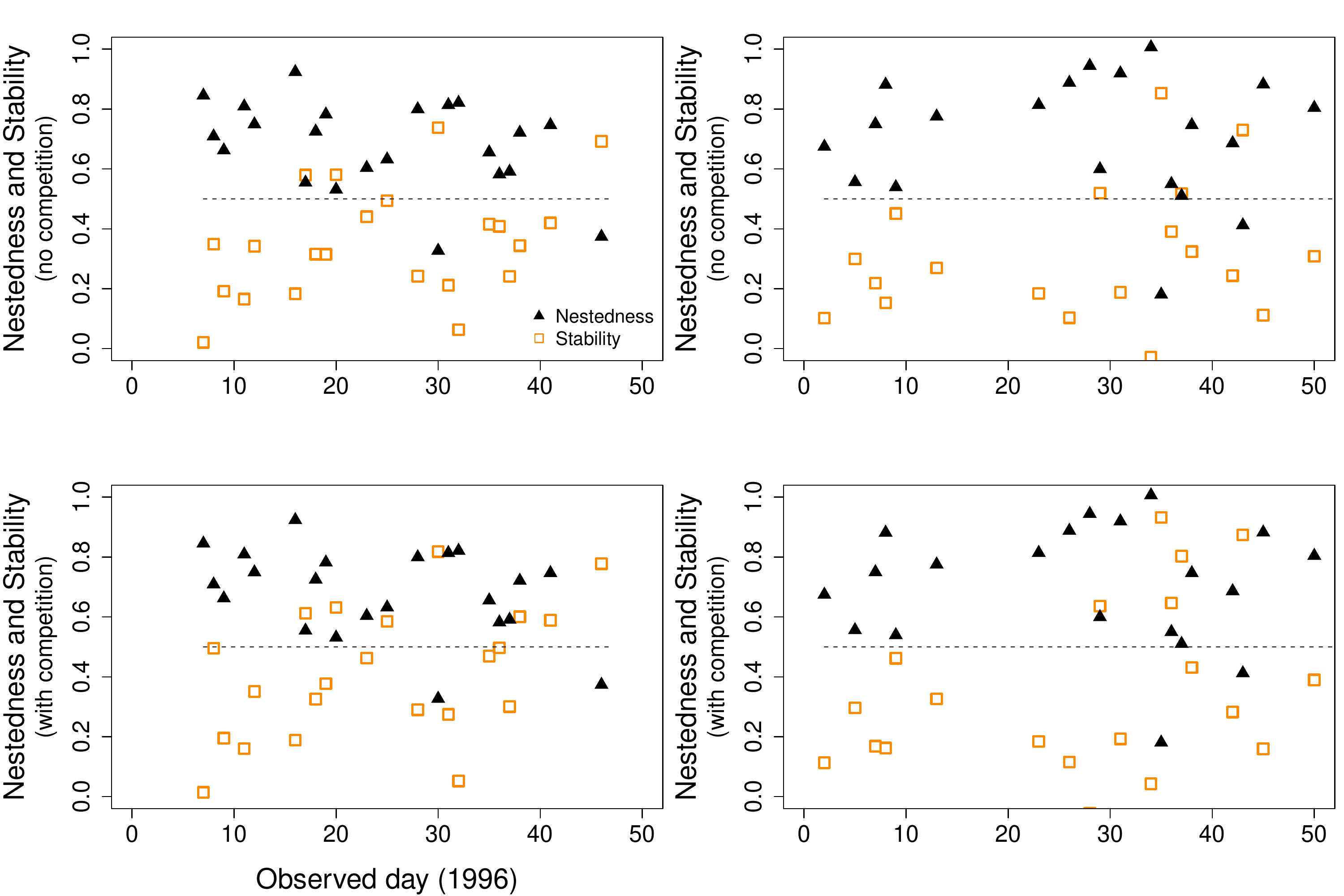}}
	\caption{Species interactions promote feasibility via nested species interaction at the expense of stability. The figure shows the scaled stability values (open squares) and the scaled nestedness values (closed triangles) in each of the observed days across the two periods (1996 and 1997). Each symbol corresponds to the scaled value in a day, and represents the position of the empirical network within the range of values generated from the rewired networks $g$. The top panels correspond to the scenario with no interspecific competition ($\rho=0$ and $\delta=0.25$), and the bottom panels correspond to the scenario with interspecific competition ($\rho=0.01$ and $\delta=0$). All the other explored combinations of parameter values yield the same qualitative results.}
\label{fig6}
\end{figure}

\end{linenumbers}
\end{spacing}
\end{document}